\documentclass[sigconf]{acmart}

\usepackage{enumitem}

\AtBeginDocument{%
  }

\copyrightyear{2026}
\acmYear{2026}
\setcopyright{cc}
\setcctype{by}
\acmConference[UIST Adjunct '26]{The 39th Annual ACM Symposium on User Interface Software and Technology}{November 02--05, 2026}{Detroit, MI, USA}
\acmBooktitle{The 39th Annual ACM Symposium on User Interface Software and Technology (UIST Adjunct '26), November 02--05, 2026, Detroit, MI, USA}
\acmDOI{10.1145/3830397.3841834}
\acmISBN{979-8-4007-2855-6/2026/11}

\begin{document}

\title{RecalibrateGPT: AI Fatigue Resilient Conversational Interfaces}


\author{Nikhil Wani}
\affiliation{%
  \institution{OpenThreads AI Research}
  \city{}
  \country{}
  }
\email{nikhylwani@openthreadsai.com}
\orcid{0009-0002-3944-6571}



\begin{abstract}
Large language models are powerful, but their interfaces often devolve into a \textit{type $\rightarrow$ read $\rightarrow$ retype} loop, creating conversational AI fatigue, cognitive load, and eventual task abandonment. To mitigate this, we present RecalibrateGPT, a system introducing five cross-turn operators (Anchor, Replay, Delta, Scope, and Steer) that each target a distinct fatigue type, recalibrating LLM responses through a structured panel by acting on the full conversation history with a single click. Users invoke these operators through the AssistiveButton in one of three operator palette layouts: Vertical, Arc, or Tablet. We conducted two pilot studies with the same 12 advanced LLM users. An initial formative qualitative study identifies a taxonomy of four fatigue types (retyping, scanning, decision paralysis, and context drift) and derives two design objectives for RecalibrateGPT. A follow-up quantitative evaluation finds it reduces perceived cognitive workload by half (NASA-TLX = 2.7) at high perceived usability (SUS = 86.5), suggesting AI fatigue is not just a model-quality issue but an interaction-flow cost that interfaces can remove.
\end{abstract}

\ccsdesc[500]{Human-centered computing~Natural language interfaces}
\ccsdesc[300]{Human-centered computing~Interactive systems and tools}
\ccsdesc[300]{Human-centered computing~User studies}
\ccsdesc[500]{Computing methodologies~Artificial intelligence}

\keywords{Human-AI Interaction, Conversational User Interfaces}

\begin{teaserfigure}
\centering
 \includegraphics
 [
    width=\textwidth,
    height=0.45\textheight,
    keepaspectratio
  ]{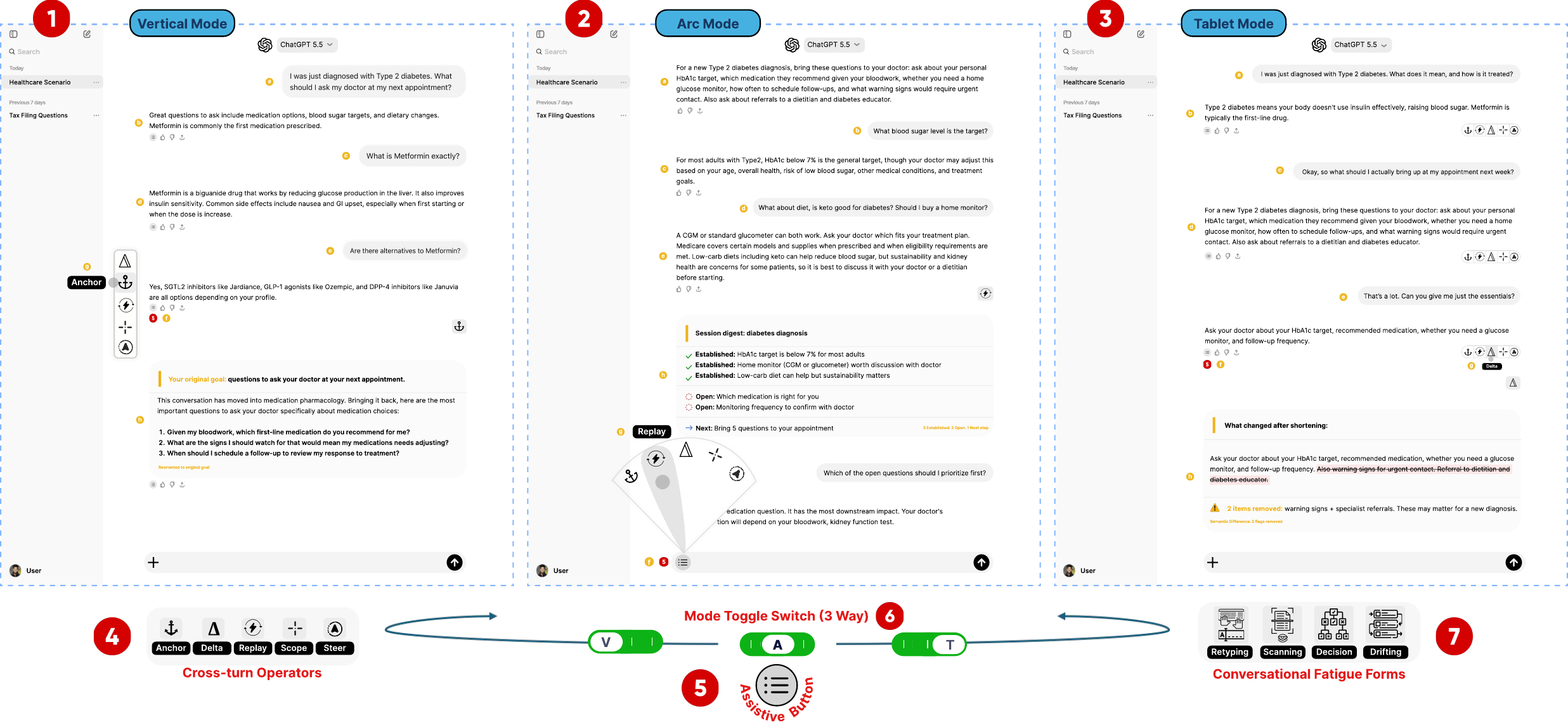}
  \caption{Three drifting healthcare conversations calibrated by RecalibrateGPT's cross-turn operators (4) across 3 palette layouts: Vertical (1), Arc (2), Tablet (3), invoked via the AssistiveButton (5) and mode-toggle (6), each targeting a fatigue form (7).}
  \Description{Three drifting healthcare conversations calibrated by RecalibrateGPT's cross-turn operators (4) across 3 palette layouts: Vertical (1), Arc (2), Tablet (3), invoked via the AssistiveButton (5) and mode-toggle (6), each targeting a fatigue form (7).}
  \label{fig:teaser}
\end{teaserfigure}

\settopmatter{printccs=true, printacmref=true}
\maketitle

\begingroup
\renewcommand{\thefootnote}{\fnsymbol{footnote}}
\footnotetext[1]{\textit{\textbf{Scope} and \textbf{Steer} operator interfaces included in supplementary material.}}
\endgroup

\section{Introduction and Related Work}
Large language models (LLMs) are powerful, but their interfaces often devolve into a \textit{type $\rightarrow$ read $\rightarrow$ retype} interaction loop: users restate goals, correct drifts, constantly keep up with relentless changes, and manage growing context across turns. Over time this produces conversational \textbf{AI fatigue}, creating cognitive load \cite{stanford-cog-challenges, metacognitive-chi24}, frustration \cite{bad-pa-chi16}, and eventual decision paralysis \cite{johy-chi23} that can end in abandonment \cite{hci-journal-discontinuance}, especially in high-stakes domains (clinical diagnosis \cite{clinical-chi24} and legal interpretation) where accuracy matters \cite{clinical_chi_21,wani-etal-2018-whole}. This fatigue is not only a model-quality issue but also an \textbf{interaction-flow issue}: many user actions currently require full re-prompting by typing, which increases per-token inference cost, compounding the burden of every additional turn.

We present \textbf{RecalibrateGPT}, a system introducing five cross-turn operators (Section~\ref{sec:system}) that each targets one of four conversational AI fatigue types (Section~\ref{sec:study1}), recalibrating LLM responses through a structured panel (Fig.~\ref{fig:teaser}(h)) by acting on the full conversation history with one click, rather than only the previous response. The AssistiveButton opens one of three operator palettes (Vertical, Arc, or Tablet) from which users select an operator; the system also calibrates interaction flow by breaking the exhausting \textit{type $\rightarrow$ read $\rightarrow$ retype} loop into a lighter, restorative \textit{type $\rightarrow$ read $\rightarrow$ click} loop, as traced in Fig.~\ref{fig:teaser}: \textit{a$\rightarrow$b$\rightarrow$c$\rightarrow$d$\rightarrow$e$\rightarrow$f$\rightarrow$g$\rightarrow$h}

\textbf{Related Work.} Prior work in UIST and the broader HCI literature explores reifying prompt actions into reusable interface controls, yet two major gaps remain: existing operators act only on a single-turn previous response, and none are designed around session-level conversational fatigue. DirectGPT \cite{directgpt-chi24} offers commands for embodying prompts as reusable direct-manipulation objects; Cells and Lenses \cite{kim-uist23-cell-generator-lens} treats LLM inputs and outputs as manipulable objects of iteration. Other systems improve single-response navigation, converting it into interactive diagrams \cite{Graphologue-uist23} or verifiable editing surfaces \cite{verifiable-uist24, sensecape-uist23}, while pipeline tools chain prompts into workflows \cite{chainforge-uist, promptchainer-chi22}. Across all of these cited prior systems and commercial interfaces (ChatGPT, Claude, Gemini), the unit of manipulation remains a single response, with fatigue unaddressed. RecalibrateGPT closes both gaps: our cross-turn operators recalibrate the entire conversation, derived from observed fatigue themes (Section~\ref{sec:study1}). To our knowledge, no prior system defines operators acting on full conversation history across three geometric palettes (Vertical, Arc, Tablet) tuned specifically to fatigue context.

\vspace{-0.6em}
\section{Study 1: Formative Study}\label{sec:study1}
To ground RecalibrateGPT in observed user interaction behavior rather than designer intuition, we conducted a qualitative formative study. \textbf{Participants.} We recruited 12 advanced LLM users (7M, 5F; ages 22–39, M=28.4, SD=4.6) through purposive sampling via social and academic networks. Inclusion criteria required daily LLM use ($\geq$ 5 days/week, $\geq$ 6 months sustained), active experience across $\geq$ 2 conversational AI platforms, English fluency (self-rated $\geq$ 4/5), and self-reported use of LLMs for high-stakes information-seeking tasks (e.g., medical or legal questions); users with exclusively single-turn, casual LLM experience were excluded. \textbf{Task.} Participants completed an online survey describing in free text their interaction frustrations and most retyped corrections during everyday LLM sessions, yielding 96 reprompt examples (M=8.0 per participant).

\textbf{Derived Design Objectives.} We derive two design objectives (DOs) grounding RecalibrateGPT. \textbf{DO1: Multi-turn response calibration}. Single-coder thematic analysis identified four recurring fatigue themes (F1–F4), each clustering around multi-turn interaction breakdown. Participants repeatedly referenced \textbf{(F1) Retyping fatigue:} "I kept retyping the same constraints because the answer drifted" [P2]; \textbf{(F2) Scanning fatigue:} "I was rereading to find the one line that actually mattered" [P11]; \textbf{(F3) Decision paralysis:} "I kept asking follow-ups because I couldn't tell what to pick" [P9]; and \textbf{(F4) Context drift:} "I was fighting the interface, not solving the problem" [P3]. These themes motivate cross-turn operators. \textbf{DO2:} \textbf{Single-click corrective control}. This motivates the AssistiveButton and three-way toggle that reduces retyping burden.

\vspace{-0.4em}
\section{RecalibrateGPT System Design}\label{sec:system}
\textbf{Cross-Turn Operators.} RecalibrateGPT's primary contribution, these five button instances (Fig.~\ref{fig:teaser}(4)) each target one of four fatigue types (F1–F4) [DO1], recalibrating LLM responses into a structured panel (Fig.~\ref{fig:teaser}(h)) by acting on the full conversation history. \textbf{Anchor} detects semantic drift between the current response and the user's original goal and reorients the output back toward it, reducing context drift (F4). \textbf{Replay} extracts a session digest of established facts, open questions, and a next step, reducing scanning fatigue (F2). \textbf{Delta (Diff)}  semantically compares the difference between current and previous response, surfacing additions and flagging goal-relevant removals, reducing retyping fatigue (F1). \textbf{Scope} narrows to a user-selected subtopic without retyping: it first surfaces detected subtopics as tappable options, then expands the selection on the second tap, reducing retyping fatigue (F1). \textbf{Steer} analyzes goal completion and generates three tappable follow-up questions targeting remaining gaps, reducing decision paralysis fatigue (F3).

\textbf{Three-Way Mode Toggle.} The toggle switch (Fig.~\ref{fig:teaser}(6)) sets the operator palette layout to one mode at a time: Vertical (V), Arc (A), or Tablet (T). Vertical renders a pill-shaped sidebar left of each response, for repeated use across turns; Arc renders a fan menu radiating from the AssistiveButton, for rapid micro-corrections while reading; Tablet renders a persistent horizontal strip below each response, for single-click access under higher cognitive load.

\textbf{Universal AssistiveButton.} The AssistiveButton (Fig.~\ref{fig:teaser}(5)) is the single entry point across all three modes for operator palette invocation. Its placement adapts (Fig.~\ref{fig:teaser}(f)): in Vertical and Tablet it appears as the first affordance after every response; in Arc it anchors to the left of the central input field. On tap, it reads the toggle switch state and renders the matching operator palette. 

\textbf{Implementation.} We implement the operator backend in Python over the OpenAI API (GPT-5.5); all operators share sentence transformer embeddings \cite{reimers-gurevych-2019-sentence} $\mathrm{Emb}(x)$. Let $H$ denote conversation history, $g$ the original goal, $c_n$ the latest response. \textbf{Anchor} computes goal alignment using cosine similarity $\cos(\mathrm{Emb}(g), \mathrm{Emb}(c_n))$ and reorients low-similarity outputs. \textbf{Delta} compares response semantic distributions using KL divergence $D_{KL}(P_n\|P_{n-1})$ and flags $g$-relevant removals. \textbf{Replay} summarizes $H \rightarrow \langle E, O, s \rangle$, where $E$ are established facts, $O$ are open questions, and $s$ is the next step. \textbf{Scope} clusters sentence embeddings $\{\mathbf{v}_{s_i} : s_i \in c_n\}$ to surface subtopics. \textbf{Steer} computes unresolved gaps as $G = \{a \in A_g : \max_{h \in H}\cos(\mathrm{Emb}(a), \mathrm{Emb}(h)) < \tau\}$, where $A_g$ are goal sub-aspects, then converts them into targeted follow-up questions. Each operator returns JSON response templates (Fig.~\ref{fig:teaser}(g)).

\begingroup
\renewcommand{\thefootnote}{\fnsymbol{footnote}}
\footnotetext[2]{\textit{\textbf{Study 2} instruments and results plots included in supplementary \textbf{appendix}.}}
\endgroup

\vspace{-0.55em}
\section{Study 2: User Feedback on RecalibrateGPT}
We conducted a follow-up within-subjects pilot with the same 12 participants. For each of cross-turn operators, participants compared the same healthcare conversation interfaces (Fig.~\ref{fig:teaser}) under standard chat and the RecalibrateGPT system, shown side-by-side in their preferred layout (Vertical n=5, Arc n=4, Tablet n=3). For each operator, they chose the less fatiguing interface and rated both on the 7-point NASA-TLX; after all five, they completed the 10-item SUS. \textbf{Findings}. Participants consistently selected RecalibrateGPT as less fatiguing, with lower perceived workload (NASA-TLX: M=2.7 vs. 5.4) and high perceived usability (SUS: M=86.5). Anchor (n=4), Replay (n=3), and Delta (n=3) were rated most useful. Given the pilot scale, we report these as directional evidence of feasibility rather than generalizable effects.

\vspace{-0.35em}
\section{Conclusion and Future Work}
We present RecalibrateGPT, a system introducing five cross-turn operators that act on the full conversation history to target conversational AI fatigue through calibrated responses and single-click interactions. In our pilot study, it reduced perceived cognitive workload by half at high perceived usability, suggesting that much of AI fatigue is not just a model-quality issue but also an interaction-flow cost that interfaces can remove. Future work will explore proactive interfaces that dynamically surface the right operator as fatigue emerges, validated through larger-scale studies.

%

\bibliographystyle{ACM-Reference-Format}
\bibliography{references}


\begin{thebibliography}{16}


\ifx \showCODEN    \undefined \def \showCODEN     #1{\unskip}     \fi
\ifx \showISBNx    \undefined \def \showISBNx     #1{\unskip}     \fi
\ifx \showISBNxiii \undefined \def \showISBNxiii  #1{\unskip}     \fi
\ifx \showISSN     \undefined \def \showISSN      #1{\unskip}     \fi
\ifx \showLCCN     \undefined \def \showLCCN      #1{\unskip}     \fi
\ifx \shownote     \undefined \def \shownote      #1{#1}          \fi
\ifx \showarticletitle \undefined \def \showarticletitle #1{#1}   \fi
\ifx \showURL      \undefined \def \showURL       {\relax}        \fi
\providecommand\bibfield[2]{#2}
\providecommand\bibinfo[2]{#2}
\providecommand\natexlab[1]{#1}
\providecommand\showeprint[2][]{arXiv:#2}

\bibitem[Arawjo et~al\mbox{.}(2023)]%
        {chainforge-uist}
\bibfield{author}{\bibinfo{person}{Ian Arawjo}, \bibinfo{person}{Priyan Vaithilingam}, \bibinfo{person}{Martin Wattenberg}, {and} \bibinfo{person}{Elena Glassman}.} \bibinfo{year}{2023}\natexlab{}.
\newblock \showarticletitle{ChainForge: An open-source visual programming environment for prompt engineering}. In \bibinfo{booktitle}{\emph{Adjunct Proceedings of the 36th Annual ACM Symposium on User Interface Software and Technology}} (San Francisco, CA, USA) \emph{(\bibinfo{series}{UIST '23 Adjunct})}. \bibinfo{publisher}{Association for Computing Machinery}, \bibinfo{address}{New York, NY, USA}, Article \bibinfo{articleno}{4}, \bibinfo{numpages}{3}~pages.
\newblock
\showISBNx{9798400700965}
\href{https://doi.org/10.1145/3586182.3616660}{doi:\nolinkurl{10.1145/3586182.3616660}}


\bibitem[Jiang et~al\mbox{.}(2023)]%
        {Graphologue-uist23}
\bibfield{author}{\bibinfo{person}{Peiling Jiang}, \bibinfo{person}{Jude Rayan}, \bibinfo{person}{Steven~P. Dow}, {and} \bibinfo{person}{Haijun Xia}.} \bibinfo{year}{2023}\natexlab{}.
\newblock \showarticletitle{Graphologue: Exploring Large Language Model Responses with Interactive Diagrams}. In \bibinfo{booktitle}{\emph{Proceedings of the 36th Annual ACM Symposium on User Interface Software and Technology}} (San Francisco, CA, USA) \emph{(\bibinfo{series}{UIST '23})}. \bibinfo{publisher}{Association for Computing Machinery}, \bibinfo{address}{New York, NY, USA}, Article \bibinfo{articleno}{3}, \bibinfo{numpages}{20}~pages.
\newblock
\showISBNx{9798400701320}
\href{https://doi.org/10.1145/3586183.3606737}{doi:\nolinkurl{10.1145/3586183.3606737}}


\bibitem[Kim et~al\mbox{.}(2023)]%
        {kim-uist23-cell-generator-lens}
\bibfield{author}{\bibinfo{person}{Tae~Soo Kim}, \bibinfo{person}{Yoonjoo Lee}, \bibinfo{person}{Minsuk Chang}, {and} \bibinfo{person}{Juho Kim}.} \bibinfo{year}{2023}\natexlab{}.
\newblock \showarticletitle{Cells, Generators, and Lenses: Design Framework for Object-Oriented Interaction with Large Language Models}. In \bibinfo{booktitle}{\emph{Proceedings of the 36th Annual ACM Symposium on User Interface Software and Technology}} (San Francisco, CA, USA) \emph{(\bibinfo{series}{UIST '23})}. \bibinfo{publisher}{Association for Computing Machinery}, \bibinfo{address}{New York, NY, USA}, Article \bibinfo{articleno}{4}, \bibinfo{numpages}{18}~pages.
\newblock
\showISBNx{9798400701320}
\href{https://doi.org/10.1145/3586183.3606833}{doi:\nolinkurl{10.1145/3586183.3606833}}


\bibitem[Laban et~al\mbox{.}(2024)]%
        {verifiable-uist24}
\bibfield{author}{\bibinfo{person}{Philippe Laban}, \bibinfo{person}{Jesse Vig}, \bibinfo{person}{Marti Hearst}, \bibinfo{person}{Caiming Xiong}, {and} \bibinfo{person}{Chien-Sheng Wu}.} \bibinfo{year}{2024}\natexlab{}.
\newblock \showarticletitle{Beyond the Chat: Executable and Verifiable Text-Editing with LLMs}. In \bibinfo{booktitle}{\emph{Proceedings of the 37th Annual ACM Symposium on User Interface Software and Technology}} (Pittsburgh, PA, USA) \emph{(\bibinfo{series}{UIST '24})}. \bibinfo{publisher}{Association for Computing Machinery}, \bibinfo{address}{New York, NY, USA}, Article \bibinfo{articleno}{20}, \bibinfo{numpages}{23}~pages.
\newblock
\showISBNx{9798400706288}
\href{https://doi.org/10.1145/3654777.3676419}{doi:\nolinkurl{10.1145/3654777.3676419}}


\bibitem[Luger and Sellen(2016)]%
        {bad-pa-chi16}
\bibfield{author}{\bibinfo{person}{Ewa Luger} {and} \bibinfo{person}{Abigail Sellen}.} \bibinfo{year}{2016}\natexlab{}.
\newblock \showarticletitle{"Like Having a Really Bad PA": The Gulf between User Expectation and Experience of Conversational Agents}. In \bibinfo{booktitle}{\emph{Proceedings of the 2016 CHI Conference on Human Factors in Computing Systems}} (San Jose, California, USA) \emph{(\bibinfo{series}{CHI '16})}. \bibinfo{publisher}{Association for Computing Machinery}, \bibinfo{address}{New York, NY, USA}, \bibinfo{pages}{5286–5297}.
\newblock
\showISBNx{9781450333627}
\href{https://doi.org/10.1145/2858036.2858288}{doi:\nolinkurl{10.1145/2858036.2858288}}


\bibitem[Masson et~al\mbox{.}(2024)]%
        {directgpt-chi24}
\bibfield{author}{\bibinfo{person}{Damien Masson}, \bibinfo{person}{Sylvain Malacria}, \bibinfo{person}{G\'{e}ry Casiez}, {and} \bibinfo{person}{Daniel Vogel}.} \bibinfo{year}{2024}\natexlab{}.
\newblock \showarticletitle{DirectGPT: A Direct Manipulation Interface to Interact with Large Language Models}. In \bibinfo{booktitle}{\emph{Proceedings of the 2024 CHI Conference on Human Factors in Computing Systems}} (Honolulu, HI, USA) \emph{(\bibinfo{series}{CHI '24})}. \bibinfo{publisher}{Association for Computing Machinery}, \bibinfo{address}{New York, NY, USA}, Article \bibinfo{articleno}{975}, \bibinfo{numpages}{16}~pages.
\newblock
\showISBNx{9798400703300}
\href{https://doi.org/10.1145/3613904.3642462}{doi:\nolinkurl{10.1145/3613904.3642462}}


\bibitem[Rajashekar et~al\mbox{.}(2024)]%
        {clinical-chi24}
\bibfield{author}{\bibinfo{person}{Niroop~Channa Rajashekar}, \bibinfo{person}{Yeo~Eun Shin}, \bibinfo{person}{Yuan Pu}, \bibinfo{person}{Sunny Chung}, \bibinfo{person}{Kisung You}, \bibinfo{person}{Mauro Giuffre}, \bibinfo{person}{Colleen~E Chan}, \bibinfo{person}{Theo Saarinen}, \bibinfo{person}{Allen Hsiao}, \bibinfo{person}{Jasjeet Sekhon}, \bibinfo{person}{Ambrose~H Wong}, \bibinfo{person}{Leigh~V Evans}, \bibinfo{person}{Rene~F. Kizilcec}, \bibinfo{person}{Loren Laine}, \bibinfo{person}{Terika Mccall}, {and} \bibinfo{person}{Dennis Shung}.} \bibinfo{year}{2024}\natexlab{}.
\newblock \showarticletitle{Human-Algorithmic Interaction Using a Large Language Model-Augmented Artificial Intelligence Clinical Decision Support System}. In \bibinfo{booktitle}{\emph{Proceedings of the 2024 CHI Conference on Human Factors in Computing Systems}} (Honolulu, HI, USA) \emph{(\bibinfo{series}{CHI '24})}. \bibinfo{publisher}{Association for Computing Machinery}, \bibinfo{address}{New York, NY, USA}, Article \bibinfo{articleno}{442}, \bibinfo{numpages}{20}~pages.
\newblock
\showISBNx{9798400703300}
\href{https://doi.org/10.1145/3613904.3642024}{doi:\nolinkurl{10.1145/3613904.3642024}}


\bibitem[Reimers and Gurevych(2019)]%
        {reimers-gurevych-2019-sentence}
\bibfield{author}{\bibinfo{person}{Nils Reimers} {and} \bibinfo{person}{Iryna Gurevych}.} \bibinfo{year}{2019}\natexlab{}.
\newblock \showarticletitle{Sentence-{BERT}: Sentence Embeddings using {S}iamese {BERT}-Networks}. In \bibinfo{booktitle}{\emph{Proceedings of the 2019 Conference on Empirical Methods in Natural Language Processing and the 9th International Joint Conference on Natural Language Processing (EMNLP-IJCNLP)}}, \bibfield{editor}{\bibinfo{person}{Kentaro Inui}, \bibinfo{person}{Jing Jiang}, \bibinfo{person}{Vincent Ng}, {and} \bibinfo{person}{Xiaojun Wan}} (Eds.). \bibinfo{publisher}{Association for Computational Linguistics}, \bibinfo{address}{Hong Kong, China}, \bibinfo{pages}{3982--3992}.
\newblock
\href{https://doi.org/10.18653/v1/D19-1410}{doi:\nolinkurl{10.18653/v1/D19-1410}}


\bibitem[Subramonyam et~al\mbox{.}(2024)]%
        {stanford-cog-challenges}
\bibfield{author}{\bibinfo{person}{Hari Subramonyam}, \bibinfo{person}{Roy Pea}, \bibinfo{person}{Christopher Pondoc}, \bibinfo{person}{Maneesh Agrawala}, {and} \bibinfo{person}{Colleen Seifert}.} \bibinfo{year}{2024}\natexlab{}.
\newblock \showarticletitle{Bridging the Gulf of Envisioning: Cognitive Challenges in Prompt Based Interactions with LLMs}. In \bibinfo{booktitle}{\emph{Proceedings of the 2024 CHI Conference on Human Factors in Computing Systems}} (Honolulu, HI, USA) \emph{(\bibinfo{series}{CHI '24})}. \bibinfo{publisher}{Association for Computing Machinery}, \bibinfo{address}{New York, NY, USA}, Article \bibinfo{articleno}{1039}, \bibinfo{numpages}{19}~pages.
\newblock
\showISBNx{9798400703300}
\href{https://doi.org/10.1145/3613904.3642754}{doi:\nolinkurl{10.1145/3613904.3642754}}


\bibitem[Suh et~al\mbox{.}(2023)]%
        {sensecape-uist23}
\bibfield{author}{\bibinfo{person}{Sangho Suh}, \bibinfo{person}{Bryan Min}, \bibinfo{person}{Srishti Palani}, {and} \bibinfo{person}{Haijun Xia}.} \bibinfo{year}{2023}\natexlab{}.
\newblock \showarticletitle{Sensecape: Enabling Multilevel Exploration and Sensemaking with Large Language Models}. In \bibinfo{booktitle}{\emph{Proceedings of the 36th Annual ACM Symposium on User Interface Software and Technology}} (San Francisco, CA, USA) \emph{(\bibinfo{series}{UIST '23})}. \bibinfo{publisher}{Association for Computing Machinery}, \bibinfo{address}{New York, NY, USA}, Article \bibinfo{articleno}{1}, \bibinfo{numpages}{18}~pages.
\newblock
\showISBNx{9798400701320}
\href{https://doi.org/10.1145/3586183.3606756}{doi:\nolinkurl{10.1145/3586183.3606756}}


\bibitem[Tankelevitch et~al\mbox{.}(2024)]%
        {metacognitive-chi24}
\bibfield{author}{\bibinfo{person}{Lev Tankelevitch}, \bibinfo{person}{Viktor Kewenig}, \bibinfo{person}{Auste Simkute}, \bibinfo{person}{Ava~Elizabeth Scott}, \bibinfo{person}{Advait Sarkar}, \bibinfo{person}{Abigail Sellen}, {and} \bibinfo{person}{Sean Rintel}.} \bibinfo{year}{2024}\natexlab{}.
\newblock \showarticletitle{The Metacognitive Demands and Opportunities of Generative AI}. In \bibinfo{booktitle}{\emph{Proceedings of the 2024 CHI Conference on Human Factors in Computing Systems}} (Honolulu, HI, USA) \emph{(\bibinfo{series}{CHI '24})}. \bibinfo{publisher}{Association for Computing Machinery}, \bibinfo{address}{New York, NY, USA}, Article \bibinfo{articleno}{680}, \bibinfo{numpages}{24}~pages.
\newblock
\showISBNx{9798400703300}
\href{https://doi.org/10.1145/3613904.3642902}{doi:\nolinkurl{10.1145/3613904.3642902}}


\bibitem[Wang et~al\mbox{.}(2021)]%
        {clinical_chi_21}
\bibfield{author}{\bibinfo{person}{Dakuo Wang}, \bibinfo{person}{Liuping Wang}, \bibinfo{person}{Zhan Zhang}, \bibinfo{person}{Ding Wang}, \bibinfo{person}{Haiyi Zhu}, \bibinfo{person}{Yvonne Gao}, \bibinfo{person}{Xiangmin Fan}, {and} \bibinfo{person}{Feng Tian}.} \bibinfo{year}{2021}\natexlab{}.
\newblock \showarticletitle{“Brilliant AI Doctor” in Rural Clinics: Challenges in AI-Powered Clinical Decision Support System Deployment}. In \bibinfo{booktitle}{\emph{Proceedings of the 2021 CHI Conference on Human Factors in Computing Systems}} (Yokohama, Japan) \emph{(\bibinfo{series}{CHI '21})}. \bibinfo{publisher}{Association for Computing Machinery}, \bibinfo{address}{New York, NY, USA}, Article \bibinfo{articleno}{697}, \bibinfo{numpages}{18}~pages.
\newblock
\showISBNx{9781450380966}
\href{https://doi.org/10.1145/3411764.3445432}{doi:\nolinkurl{10.1145/3411764.3445432}}


\bibitem[Wani et~al\mbox{.}(2018)]%
        {wani-etal-2018-whole}
\bibfield{author}{\bibinfo{person}{Nikhil Wani}, \bibinfo{person}{Sandeep Mathias}, \bibinfo{person}{Jayashree~Aanand Gajjam}, {and} \bibinfo{person}{Pushpak Bhattacharyya}.} \bibinfo{year}{2018}\natexlab{}.
\newblock \showarticletitle{The Whole is Greater than the Sum of its Parts: Towards the Effectiveness of Voting Ensemble Classifiers for Complex Word Identification}. In \bibinfo{booktitle}{\emph{Proceedings of the Thirteenth Workshop on Innovative Use of {NLP} for Building Educational Applications}}, \bibfield{editor}{\bibinfo{person}{Joel Tetreault}, \bibinfo{person}{Jill Burstein}, \bibinfo{person}{Ekaterina Kochmar}, \bibinfo{person}{Claudia Leacock}, {and} \bibinfo{person}{Helen Yannakoudakis}} (Eds.). \bibinfo{publisher}{Association for Computational Linguistics}, \bibinfo{address}{New Orleans, Louisiana}, \bibinfo{pages}{200--205}.
\newblock
\href{https://doi.org/10.18653/v1/W18-0522}{doi:\nolinkurl{10.18653/v1/W18-0522}}


\bibitem[Wu et~al\mbox{.}(2026)]%
        {hci-journal-discontinuance}
\bibfield{author}{\bibinfo{person}{Siyuan Wu}, \bibinfo{person}{Guochao Peng}, \bibinfo{person}{Shuyang Li}, \bibinfo{person}{David Cameron}, \bibinfo{person}{Jun Zhang}, \bibinfo{person}{Zuopeng Zhang}, {and} \bibinfo{person}{Qing Zhang}.} \bibinfo{year}{2026}\natexlab{}.
\newblock \showarticletitle{Textual cues, cognitive load, and social fatigue: Unveiling the reasons behind user discontinuance in conversational AI}.
\newblock \bibinfo{journal}{\emph{International Journal of Human-Computer Studies}}  \bibinfo{volume}{215} (\bibinfo{year}{2026}), \bibinfo{pages}{103846}.
\newblock
\showISSN{1071-5819}
\href{https://doi.org/10.1016/j.ijhcs.2026.103846}{doi:\nolinkurl{10.1016/j.ijhcs.2026.103846}}


\bibitem[Wu et~al\mbox{.}(2022)]%
        {promptchainer-chi22}
\bibfield{author}{\bibinfo{person}{Tongshuang Wu}, \bibinfo{person}{Ellen Jiang}, \bibinfo{person}{Aaron Donsbach}, \bibinfo{person}{Jeff Gray}, \bibinfo{person}{Alejandra Molina}, \bibinfo{person}{Michael Terry}, {and} \bibinfo{person}{Carrie~J Cai}.} \bibinfo{year}{2022}\natexlab{}.
\newblock \showarticletitle{PromptChainer: Chaining Large Language Model Prompts through Visual Programming}. In \bibinfo{booktitle}{\emph{Extended Abstracts of the 2022 CHI Conference on Human Factors in Computing Systems}} (New Orleans, LA, USA) \emph{(\bibinfo{series}{CHI EA '22})}. \bibinfo{publisher}{Association for Computing Machinery}, \bibinfo{address}{New York, NY, USA}, Article \bibinfo{articleno}{359}, \bibinfo{numpages}{10}~pages.
\newblock
\showISBNx{9781450391566}
\href{https://doi.org/10.1145/3491101.3519729}{doi:\nolinkurl{10.1145/3491101.3519729}}


\bibitem[Zamfirescu-Pereira et~al\mbox{.}(2023)]%
        {johy-chi23}
\bibfield{author}{\bibinfo{person}{J.D. Zamfirescu-Pereira}, \bibinfo{person}{Richmond~Y. Wong}, \bibinfo{person}{Bjoern Hartmann}, {and} \bibinfo{person}{Qian Yang}.} \bibinfo{year}{2023}\natexlab{}.
\newblock \showarticletitle{Why Johnny Can’t Prompt: How Non-AI Experts Try (and Fail) to Design LLM Prompts}. In \bibinfo{booktitle}{\emph{Proceedings of the 2023 CHI Conference on Human Factors in Computing Systems}} (Hamburg, Germany) \emph{(\bibinfo{series}{CHI '23})}. \bibinfo{publisher}{Association for Computing Machinery}, \bibinfo{address}{New York, NY, USA}, Article \bibinfo{articleno}{437}, \bibinfo{numpages}{21}~pages.
\newblock
\showISBNx{9781450394215}
\href{https://doi.org/10.1145/3544548.3581388}{doi:\nolinkurl{10.1145/3544548.3581388}}


\end{thebibliography}


\setcounter{figure}{0}
\renewcommand{\thefigure}{\Alph{figure}}

\raggedbottom

\twocolumn[{%
  \section*{A. Appendix}
  \vspace{10pt}
  \begin{center}
    \includegraphics[
      width=\textwidth,
      height=0.39\textheight,
      keepaspectratio
    ]{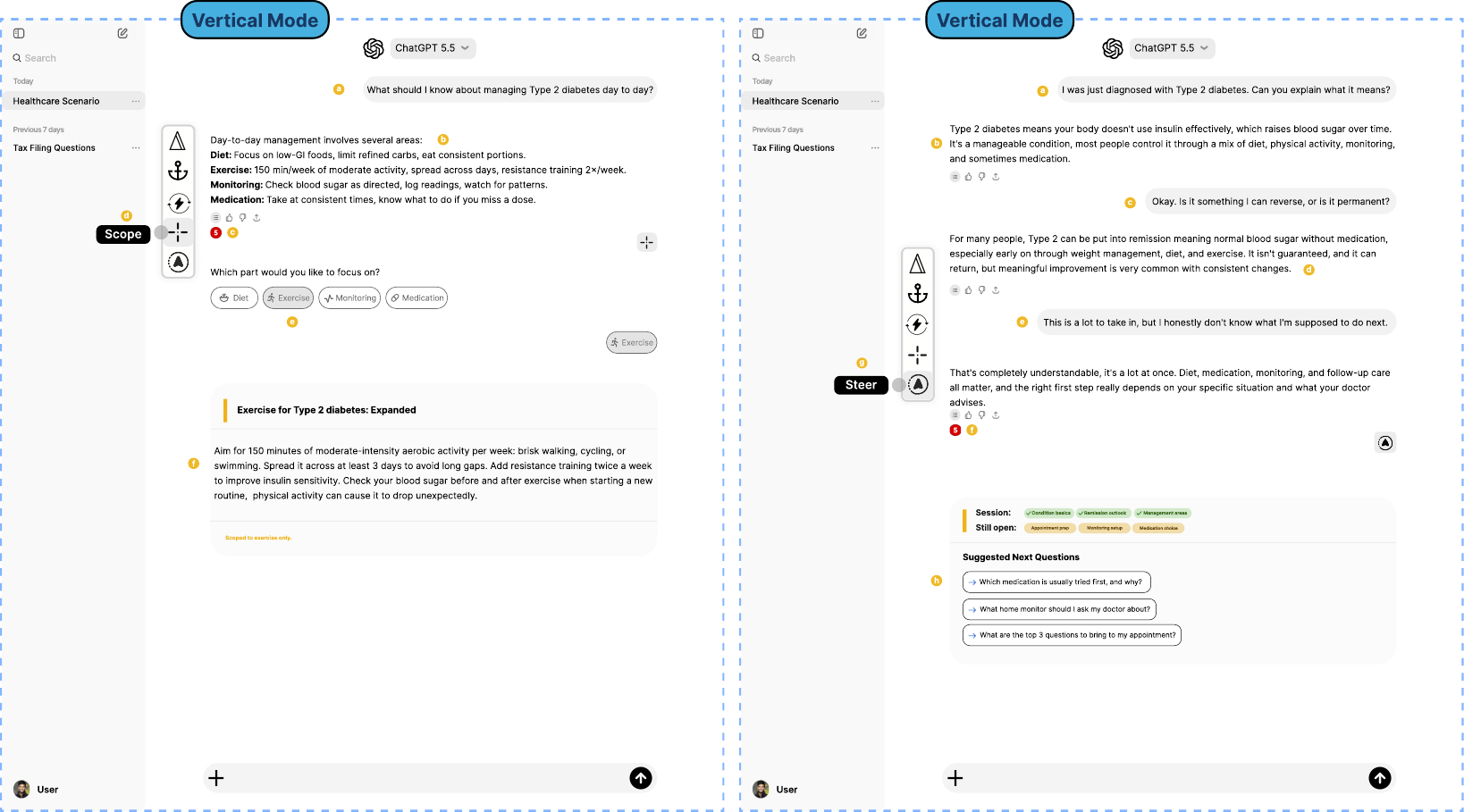}
    \captionof{figure}{Two drifting healthcare conversations calibrated by RecalibrateGPT’s Scope (left) and Steer (right) cross-turn operator interfaces in the Vertical palette layout.}
    \Description{Two drifting healthcare conversations calibrated by RecalibrateGPT’s Scope (left) and Steer (right) cross-turn operator interfaces in the Vertical palette layout.}
    \label{fig:scope-steer}
  \end{center}
  \vspace{13pt}
}]

\noindent This appendix details the Study 2 procedure, participant feedback and measurement instruments. Results are reported in the main paper; the summary below, Fig.~\ref{fig:scope-steer}, and Fig.~\ref{fig:study2} are provided for completeness.

\subsection*{A.1 Study 2 Procedure}
Sessions were conducted online, approximately 35 minutes per participant, across three phases. Participants were compensated for their time. The study protocol was reviewed and approved by an internal ethics review process. All participants provided informed consent.

\vspace{0.4em}
\noindent\textbf{Phase 1: Onboarding (5 min).} Participants were briefed on the objective, re-verified against inclusion criteria, and introduced to the scenario.

\begin{figure*}[t]
  \centering
  \includegraphics[
    width=\textwidth,
    height=0.42\textheight,
    keepaspectratio
  ]{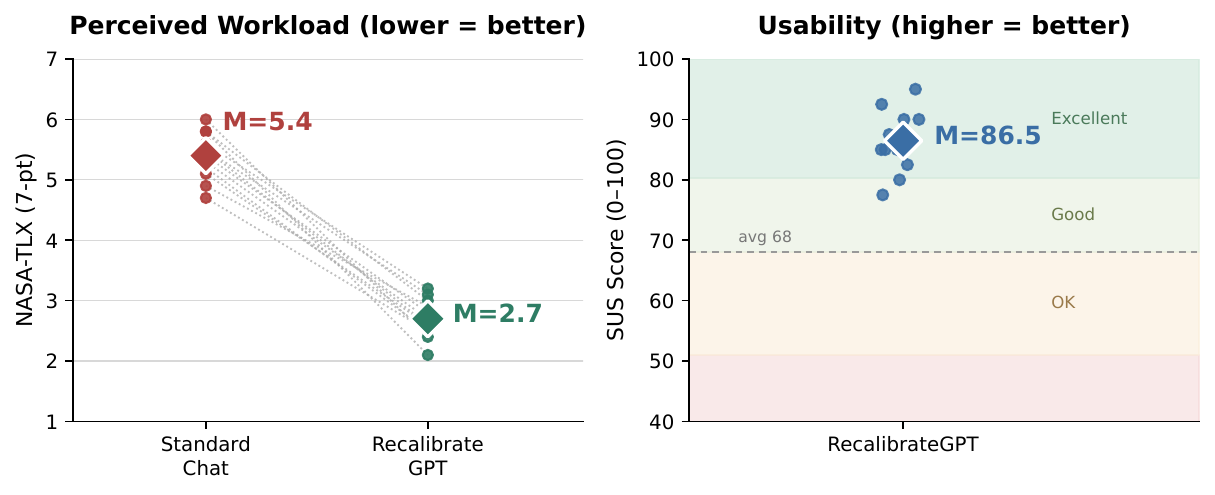}
  \caption{Study 2 participant feedback (N=12). Left: participant-level mean workload (NASA-TLX) ratings were lower (M = 2.7 vs. 5.4) under RecalibrateGPT. Right: SUS scores (M=86.5) fall in the ``excellent'' usability band.}
  \Description{Two panels. Left: a paired-slopes plot of participant-level mean NASA-TLX workload, dropping from 5.4 under standard chat to 2.7 under RecalibrateGPT for all 12 participants. Right: a dot plot of SUS scores clustered around a mean of 86.5, within the excellent usability band above 80.}
  \label{fig:study2}
\end{figure*}

\vspace{0.4em}
\noindent\textbf{Phase 2: Paired Evaluation (20 min).} We employed a within-subjects design (N=12). Participants evaluated a simulated healthcare scenario, shown in Fig.~\ref{fig:teaser} of the main paper, across five cross-turn operator-specific comparisons involving the management of a new Type 2 diabetes diagnosis. For each operator (Anchor, Replay, Delta, Scope, and Steer), they viewed a side-by-side pair of conversation interfaces: Condition A, a standard-chat baseline, and Condition B, a RecalibrateGPT intervention, resulting in 10 screens total. Each pair was shown together in the participant's preferred layout (Vertical N=5, Arc N=4, Tablet N=3), with operator order and the presentation order of the two conditions randomized per participant to reduce order effects. After each comparison, participants selected the less fatigue-inducing interface and completed parallel 7-point NASA-TLX ratings for both conditions. For each operator-condition comparison, the six NASA-TLX dimension ratings were averaged to obtain an unweighted NASA-TLX score. The five operator scores were then averaged to obtain one participant-level mean per condition. Fig.~\ref{fig:study2} plots these paired means.

\vspace{0.4em}
\noindent\textbf{Phase 3: Usability and Debrief (10 min).} After all five pair comparisons, participants rated the RecalibrateGPT panel on the 10-item SUS, followed by a brief semi-structured interview to understand which elements affected retyping, scanning, decision-making, and context maintenance.

\subsection*{A.2 Participant Feedback}
\noindent\textbf{Observations.} Participants uniformly selected RecalibrateGPT as the less fatigue-inducing interface, reported lower perceived workload (NASA-TLX: M=2.7 vs.\ 5.4), and rated it highly usable (SUS: M=86.5); Anchor (n=4), Replay (n=3), and Delta (n=3) were most frequently identified as useful during debrief.

\subsection*{A.3 Measurement Instruments}

\noindent\textbf{1. NASA-TLX (Unweighted, 7-point).}
Participants rated each interface on the six standard NASA-TLX dimensions, each on a 7-point scale (1 = Very Low, 7 = Very High); Performance used reversed anchors from 1 (Very Successful) to 7 (Very Unsuccessful), so that higher values consistently indicated greater perceived workload:
\begin{enumerate}[leftmargin=*, labelindent=0pt, labelsep=0.5em, itemsep=2pt, topsep=4pt, parsep=0pt]
    \item \textbf{Mental Demand}: How mentally demanding was the interface?
    \item \textbf{Physical Demand}: How physically demanding was the interface?
    \item \textbf{Temporal Demand}: How hurried or rushed was the pace of the task (conversational interaction flow)?
    \item \textbf{Performance}: How successful were you in accomplishing what you were asked to do?
    \item \textbf{Effort}: How hard did you have to work to accomplish your level of performance?
    \item \textbf{Frustration}: How insecure, discouraged, irritated, stressed, and annoyed were you?
\end{enumerate}

\vspace{0.4em}
\noindent\textbf{2. System Usability Scale (SUS), 10-item, 5-point Likert}
(1 = Strongly Disagree, 5 = Strongly Agree), scored 0--100:

\begin{enumerate}[leftmargin=*, labelindent=0pt, labelsep=0.5em, itemsep=2pt, topsep=4pt, parsep=0pt]
    \item I think that I would like to use this system frequently.
    \item I found the system unnecessarily complex.
    \item I thought the system was easy to use.
    \item I think that I would need the support of a technical person to be able to use this system.
    \item I found the various functions in this system were well integrated.
    \item I thought there was too much inconsistency in this system.
    \item I would imagine that most people would learn to use this system very quickly.
    \item I found the system very cumbersome to use.
    \item I felt very confident using the system.
    \item I needed to learn a lot of things before I could get going with this system.
\end{enumerate}

\end{document}